\documentclass[superscriptaddress, reprint, amsmath, amssymb,
aps,
pra,
floatfix]{revtex4-2}

\usepackage{mathptmx}
\usepackage{xcolor}
\usepackage{hyperref}
\hypersetup{
  colorlinks=true,
  linkcolor=blue,
  filecolor=magenta,
  urlcolor=magenta,
}

\usepackage{graphicx}
\usepackage{subfigure}
\usepackage{multirow}
\usepackage{orcidlink}
\usepackage{float}
\usepackage{bm}
\usepackage{amsmath, amssymb}

\begin{document}
\title{The Particle in a Box in Koopman--von Neumann Mechanics: A Hilbert Space representation of Classical Mechanics}

\author{Abhijit Sen \orcidlink{0000-0003-2783-1763}}
\email{abhijit913@gmail.com}
\affiliation{Department of Physics and Engineering Physics, Tulane University, New Orleans, Louisiana 70118, USA}

\author{Lev Kaplan\orcidlink{0000-0002-7256-3203}}
\email{lkaplan@tulane.edu}
\affiliation{Department of Physics and Engineering Physics, Tulane University,  New Orleans, Louisiana 70118, USA}

\begin{abstract}
This paper revisits the textbook 'particle in a box', but from the point of view of Koopman-von Neumann (KvN) mechanics. KvN mechanics is a way to describe \emph{classical} dynamics in a Hilbert space.  That simple fact changes the usual expectation: hard walls do \emph{not} force energy quantization here. We show, in a clear and physical way, why a KvN particle confined between two ideal walls still has a continuous range of energies. With the correct wall condition, one that captures ordinary elastic reflection rather than 'vanishing at the boundary,' the KvN description naturally produces spatial confinement without discrete energy levels. Beyond establishing this result, we also clean up common misunderstandings: for example, treating the KvN wavefunction like a quantum probability amplitude in position alone leads to the wrong boundary picture and, with it, the wrong conclusion about quantization.

\end{abstract}

\maketitle

\section{Introduction}

The relationship between classical and quantum mechanical descriptions of physical systems remains a fundamental question in theoretical physics. While classical mechanics provides a deterministic framework based on trajectories in phase space, quantum mechanics operates within the structure of Hilbert spaces, where boundary conditions can lead to energy quantization and discrete spectra, and measurement outcomes are probabilistic. The transition between these frameworks and the role of mathematical formalism versus physical content continues to generate important insights.

The Koopman-von Neumann (KvN) approach~\cite{9,8,Sen2020,Sen2023,Sen_2023,Sen-2023,Bondar_2012a,MAURO2002,GOZZI2004,Klein2017,Joseph2020,Deotto2003, Deotto22003} offers a unique perspective on this classical-quantum relationship by recasting classical mechanics in the mathematical language of Hilbert spaces. In this formulation, classical evolution is described by a complex wavefunction on phase space that evolves unitarily under the classical Hamiltonian flow. Crucially, this mathematical restructuring preserves the underlying classical physics while adopting the linear algebraic framework typically associated with quantum theory~\cite{Mauro:2001rm,Mauro}.

A paradigmatic example for understanding quantum-classical distinctions is the particle confined to a one-dimensional box with infinite potential walls. In quantum mechanics, the boundary conditions imposed by perfectly reflecting walls lead directly to energy eigenvalue quantization and the familiar discrete spectrum $E_n = n^2\pi^2\hbar^2/(2mL^2)$~\cite{Shankar_1994}. This quantization arises from the requirement that the wavefunction vanish at the boundaries, constraining the allowed wavelengths to discrete values.

The question naturally arises: does spatial confinement necessarily imply energy quantization when classical mechanics is formulated in Hilbert space? This investigation addresses a fundamental issue regarding which aspects of discrete spectra stem from quantum mechanical principles versus which arise merely from the mathematical framework of Hilbert spaces~\cite{Klein_2018}.

In this work, we examine the classical particle confined between perfectly elastic walls using the KvN formalism. We demonstrate that while spatial confinement constrains the functional form of the KvN wavefunction, it does not lead to energy quantization. The appropriate boundary condition corresponds to ideal elastic reflection, preserving probability current and maintaining unitarity of the evolution. The resulting continuous energy spectrum illustrates that discretization is a genuinely quantum phenomenon rather than a consequence of Hilbert space formulation alone.

This analysis contributes to the broader understanding of classical-quantum correspondence by isolating the mathematical aspects of Hilbert space methods from the physical content that distinguishes quantum mechanics~\cite{Giulini:1996nw,Klein2012}. Our results clarify the distinct roles of boundary conditions in classical versus quantum theories and provide insight into the emergence of discrete spectra in quantum systems.

\section{Koopman--von Neumann mechanics: A Hilbert space formulation of classical dynamics}
\label{sec:kvn-formalism}

The KvN approach  ~\cite{9,8} formulates classical mechanics on a Hilbert space by evolving a complex phase-space amplitude with the Liouville (Koopman) generator. The evolution is unitary, yet the basic observables \(\hat q\) and \(\hat p\) commute, so the underlying physics remains purely classical. This framework imports linear, operator-theoretic tools without quantum-specific features such as noncommutativity or discrete, boundary-induced spectra ~\cite{Mauro:2001rm,Mauro}.

Concretely, if $\rho(q,p,t)$ is the Liouville density on phase space, one introduces a complex ``classical wavefunction'' $\psi(q,p,t)$ such that $|\psi|^2=\rho$ and demands that it obey the Liouville equation in Schr\"odinger form
\begin{equation}
i\hbar\,\partial_t \psi(q,p,t)\;=\;\hat L\,\psi(q,p,t), 
\qquad 
\hat L \;=\; i\hbar\Big(\partial_q H\,\partial_p-\partial_p H\,\partial_q\Big),
\label{eq:Liouville-Schro}
\end{equation}
with the usual classical Hamiltonian $H(q,p,t)$. The evolution is unitary with respect to the inner product
\begin{equation}
\langle \psi|\phi\rangle \;=\;\int dq\,dp\;\psi^*(q,p,t)\,\phi(q,p,t),
\end{equation}
and expectation values are computed by $\langle \Lambda \rangle = \int dq\,dp\;\psi^*\,\hat\Lambda\,\psi$. A crucial structural difference from quantum mechanics is that, in KvN, the multiplicative operators $\hat q$ and $\hat p$ commute, $\,[\hat q,\hat p]=0$, and act as $\hat q\,\psi=q\psi$, $\hat p\,\psi=p\psi$; the generator $\hat L$ is first order in derivatives~\cite{Gozzi:2003sh}.

Two complementary pictures are useful. In the Heisenberg picture with $\hat\Lambda(t)=e^{i\hat L t}\hat\Lambda(0)e^{-i\hat L t}$ one finds
\begin{equation}
\frac{d\hat q}{dt}= i[\hat L,\hat q]=\partial_p H,\qquad 
\frac{d\hat p}{dt}= i[\hat L,\hat p]=-\,\partial_q H,
\end{equation}
which are just Hamilton's equations. In the Schr\"odinger picture, the method of characteristics shows that $\psi$ is transported along classical trajectories in $(q,p)$ generated by $H$~\cite{wilczek_2022}.

Sudarshan proposed viewing classical mechanics as a special sector of a larger quantum system with \emph{doubled} canonical pairs~\cite{Sudarshan_1976,Sherry1978}. Introduce operators $(\hat q,\hat p)$ and a second pair $(\hat Q,\hat P)$ so that the nonvanishing commutators are
\begin{equation}
[\hat q,\hat P]=i\hbar,\qquad [\hat Q,\hat p]=i\hbar,
\end{equation}
with all others (including $[\hat q,\hat p]$) vanishing. In this extended space, the KvN generator becomes the genuinely quantum Hamiltonian
\begin{equation}
\hat H_c=\frac{\partial H}{\partial q}\,\hat Q+\frac{\partial H}{\partial p}\,\hat P,
\qquad
i\hbar\,\partial_t\psi=\hat H_c\,\psi,
\label{eq:KvN-extended-H}
\end{equation}
so that classical KvN evolution is unitarily generated by a linear Hamiltonian in the ``hidden'' partners $(\hat Q,\hat P)$~\cite{Bondar_2012a}.

Two representations are especially useful. In the phase-space $(q,p)$ representation, $\hat q$ and $\hat p$ act multiplicatively, while $\hat Q=i\hbar\,\partial_p$ and $\hat P=-\,i\hbar\,\partial_q$. Substituting these into \eqref{eq:KvN-extended-H} returns the Liouville operator of Eq.~\eqref{eq:Liouville-Schro}~\cite{Bondar_2013}. In the $(q,Q)$ representation, obtained by Fourier transforming in $p$:
\begin{equation}
\psi(q,Q,t)=\frac{1}{\sqrt{2\pi}}\int dp\,e^{ipQ/\hbar}\,\psi(q,p,t),
\end{equation}
the operators $\hat q$ and $\hat Q$ are multiplicative, while $\hat p=-\,i\hbar\,\partial_Q$ and $\hat P=-\,i\hbar\,\partial_q$~\cite{Hillery1984}.

For the standard Hamiltonian $H(q,p)=\frac{p^2}{2m}+V(q)$, Eq.~\eqref{eq:KvN-extended-H} becomes, explicitly in $(q,Q)$,
\begin{equation}
i\hbar\,\partial_t\psi(q,Q,t)
=
\Big[-\,\frac{\hbar^2}{m}\,\partial_q\partial_Q + V'(q)\,Q\Big]\psi(q,Q,t),
\label{eq:qQ-evolution}
\end{equation}
which highlights the ``mixed-derivative'' structure characteristic of KvN in this representation. Note that the apparent absence of the factor $\tfrac{1}{2}$ is not an algebraic oversight but a reflection of a deeper structural distinction between the Schrödinger and KvN formulations of dynamics. In quantum mechanics, the Schrödinger equation employs the Hamiltonian $\hat{H}=\hat{p}^{2}/(2m)$ as the generator of time evolution, so that the wavefunction acquires energy–dependent phases through its quadratic dependence on momentum. The evolution thus encodes interference and quantization through oscillatory phase accumulation proportional to the kinetic energy. In KvN mechanics, by contrast, time evolution is governed by the Liouvillian operator, which generates \emph{classical flow} in phase space rather than energy–phase rotation in Hilbert space. In the mixed $(q,Q)$ representation, the generator takes the form $(p/m)\,\partial_q$, describing the transport of probability amplitude along deterministic trajectories with velocity $p/m$. Because the Liouvillian depends linearly on momentum, it reflects the kinematics of motion rather than the energetics of phase, and the factor of $\tfrac{1}{2}$, characteristic of kinetic energy, naturally does not appear.

A key distinction emerges between representations: in the $(q,p)$ representation, writing $\psi=\sqrt{\rho}\,e^{iS}$ yields two \emph{decoupled} Liouville equations for $\sqrt{\rho}$ and $S$, mirroring the absence of quantum-like interference in classical evolution~\cite{Gozzi:2001he}. In $(q,Q)$, by contrast, the amplitude and phase \emph{couple} through the mixed derivatives in \eqref{eq:qQ-evolution}.

Sudarshan explained the absence of observable interference by a superselection rule tied to the classical observable set: classical observables are functions of $(q,p)$ and commute with $\hat q,\hat p$. Coherent superpositions of distinct $|q,p\rangle$ then become operationally indistinguishable from incoherent mixtures, since off-diagonal matrix elements vanish between different $|q,p\rangle$ sectors~\cite{Abrikosov:2004cf}.

\section{Koopman--von Neumann mechanics: Deeper Insight}
\label{sec:kvn}
Koopman--von Neumann (KvN) mechanics can be placed alongside the familiar Schr\"odinger$\to$Hamilton--Jacobi correspondence, but with the correct phase-space control knob ($\kappa$). In configuration space, the Madelung split $\psi=\sqrt{\rho}\,e^{iS/\hbar}$ and the limit $\hbar\to0$ yield the continuity equation for $\rho$ together with the Hamilton--Jacobi equation for $S$ ~\cite{Padmanabhan2015}, i.e., an ensemble-classical evolution rather than a single Newtonian trajectory. 
The phase-space twin of this story begins from the Wigner--Moyal formulation with a commutativity parameter $\kappa$.
For one spatial dimension, define the $\kappa$–Wigner transform by
\begin{equation}\label{eq:Wkappa}
W(q,p;t)=\frac{1}{\sqrt{2\pi\,\kappa\hbar}}
\int_{\mathbb{R}} 
e^{\,i p y/(\kappa\hbar)}\,
\Psi^{*}\!\left(q+\tfrac{y}{2},t\right)\,
\Psi\!\left(q-\tfrac{y}{2},t\right)\,dy \,,
\end{equation}
where
\begin{equation}\label{eq:comm}
[\hat q,\hat p]=i\,\kappa\hbar \,,
\end{equation}
then make the change of variables
\begin{equation}\label{eq:y_sub}
y=\kappa\hbar\,\lambda_p \,,
\end{equation}
and
\begin{equation}\label{eq:uv_def}
u=q-\frac{\kappa\hbar\,\lambda_p}{2},
\qquad
v=q+\frac{\kappa\hbar\,\lambda_p}{2},
\end{equation}
and define the two-point object
\begin{equation}\label{eq:rho_def}
\rho(u,v,t)=\Psi^{*}(v,t)\,\Psi(u,t).
\end{equation}

If $\Psi$ obeys the Schr\"odinger equation with the effective Planck constant $\kappa\hbar$,
\begin{equation}\label{eq:sch_kappa}
i\,\kappa\hbar\,\partial_t \Psi(q,t)
=\left[-\frac{(\kappa\hbar)^2}{2m}\,\partial_q^2+V(q)\right]\Psi(q,t),
\end{equation}
then $\rho$ evolves exactly according to
\begin{equation}\label{eq:rho_evo}
i\,\kappa\hbar\,\partial_t \rho(u,v,t)=\big(\hat H_u-\hat H_v\big)\,\rho(u,v,t),
\end{equation}
with
\begin{equation}\label{eq:HuHv}
\hat H_u=-\frac{(\kappa\hbar)^2}{2m}\,\partial_u^2+V(u),
\qquad
\hat H_v=-\frac{(\kappa\hbar)^2}{2m}\,\partial_v^2+V(v).
\end{equation}


The dynamics of Eq.~(\ref{eq:rho_evo}) contracts
to the KvN (Liouville) equation (\ref{eq:qQ-evolution}) in the classical limit $\kappa\to0$  (see  ~\cite{Sen2023} for a complete derivation). Thus KvN mechanics emerges from phase-space quantum mechanics by turning the \emph{commutativity} dial $\kappa\to0$, exactly as Hamilton--Jacobi emerges from Schr\"odinger by turning the \emph{quantumness} dial $\hbar\to0$. 

Consider again the following equation
\begin{equation}
i\hbar\,\partial_t\psi \;=\; \hat L\,\psi,
\qquad 
\hat L \;=\; i\hbar\,\{H,\cdot\},
\label{eq:kvn}
\end{equation}
with the Poisson-bracket Liouvillian $\hat L$. In this classical lift, the Madelung form $\psi=\sqrt{f}\,e^{iS/\hbar}$ yields \emph{decoupled} equations for $\sqrt{f}$ and $S$; the quantum-pressure coupling is absent meaning that the KvN dynamics is strictly classical.

The operator structure also signals classicality: because the Liouvillian \eqref{eq:kvn} is first order in phase-space derivatives, the physical density $f=|\psi|^2$ obeys precisely the classical Liouville equation
\begin{equation}
\partial_t f \;=\; \{H,f\},
\label{eq:liouville}
\end{equation}
and any rephasing $\psi\!\to\!e^{i\chi}\psi$ leaves all admissible (classical) observables invariant. This structural observation translates into testable discriminants. Two-slit physics in KvN mechanics is straightforward: the fringe-forming phase sits in a sector that does not enter the observable 
$q$-dynamics ($\dot{q} = p/m$)~\cite{Mauro}. The observed pattern is just the sum of contributions from the two slits, with no cross terms. Thus, the absence of interference is a direct signature of KvN’s strictly classical character, with the Hilbert-space structure serving only as linear bookkeeping for ensembles and not as a source of quantum coherence.

The Aharonov-Bohm effect disappears in KvN mechanics because gauge phases are confined to an unobservable sector ~\cite{Mauro}. Unlike quantum mechanics, where such phases alter measurable outcomes (for example, interference patterns or spectra), the KvN structure ensures that they remain dynamically inert. While these phases persist mathematically, they exert no influence on particle trajectories, energy levels, or any observable quantity, all of which evolve according to purely classical Liouville dynamics. This strict separation between hidden phases and physical observables is the hallmark of the classical nature of KvN mechanics.

Gravitational free fall provides another test that illuminates the fundamental difference between quantum and KvN mechanics. Following Sudarshan's perspective and working in double canonical pairs $(q,P;Q,p)$, the KvN propagator in a uniform field differs from the free propagator only by an $Q$-dependent phase together with the expected Galilean shift. In the $(q,Q)$ representation one may write schematically
\begin{equation}
\psi_{\mathrm{grav}}(q,Q,t) \;=\; 
\exp\!\big[-\,i\,Q\,g\,t\big]\;
\psi_{\mathrm{free}}\!\left(q+\tfrac{1}{2}g t^2,\,Q,\,t\right) \,.
\label{eq:freefall}
\end{equation}
Since this gravitational phase involves hidden variables that are fundamentally unobservable in KvN mechanics, it cannot contribute to  Colella-Overhauser-Werner (COW) interference effects ~\cite{Sen2020free}. This absence of gravitational interference, stemming directly from the phase's dependence on hidden variables, exemplifies how KvN's hidden sector structure eliminates the nonlocal coherence effects that characterize quantum behavior. 

The classical nature of KvN mechanics can also be understood from a different angle. From the Liouville (ensemble/fluid) viewpoint, KvN is a linear encoding of \eqref{eq:liouville}; with the hydrodynamic substitution
\begin{equation}
f(x,p,t) \;=\; \rho(x,t)\,\delta\!\big(p-m\,v(x,t)\big),
\label{eq:hydro}
\end{equation}
one recovers the continuity equation and the Euler/Hamilton--Jacobi momentum balance without quantum stress, and because amplitude and phase decouple in the KvN lift, phases cannot modulate observable densities ~\cite{Sen2020free}. 

Recent advances in quantum-mechanics-free subsystems (QMFS) demonstrate that KvN mechanics isn’t just theoretical; it can be engineered in real systems by doubling phase-space variables and designing Hamiltonians that confine all quantum-like phases to a hidden sector, leaving only classical behavior in observable dynamics. Using techniques such as negative-mass reference frames or coupled oscillators, researchers have built physical systems where the measurable part evolves classically (following Liouville’s equation) while surplus phases remain invisible, proving KvN’s practical relevance ~\cite{Silagadze2023}.


\section{KvN particle in an infinite square well}
A paradigmatic example for understanding quantum-classical distinctions is the particle confined to a one-dimensional box with infinite potential walls. In this section we address a natural question: when classical mechanics is cast in a Hilbert-space framework, does spatial confinement by itself entail energy quantization?

In what follows, we examine the classical particle confined between perfectly elastic walls using the KvN formalism. We demonstrate that while spatial confinement constrains the functional form of the KvN wavefunction, it does not lead to energy quantization. The appropriate boundary condition corresponds to ideal elastic reflection, preserving probability current and maintaining unitarity of the evolution. The resulting continuous energy spectrum illustrates that discretization is a genuinely quantum phenomenon rather than a consequence of Hilbert space formulation alone~\cite{Klein2012}.
Consider
\begin{equation}
  V(q)=\begin{cases}
    0, & 0<q<L,\\[2pt]
    \infty, & \text{otherwise}.
  \end{cases}
\end{equation}
Inside the box $V'(q)=0$, so Eq.~\eqref{eq:qQ-evolution} becomes
\begin{equation}
\label{eq:box-PDE}
  i\hbar\,\partial_t \psi(q,Q,t) = -\frac{\hbar^2}{m}\,\partial_q\partial_Q \psi(q,Q,t).
\end{equation}

The boundary conditions at the hard walls determine the physical content of this evolution. An ideal classical wall implements specular reflection: $(q_w,p) \mapsto (q_w,-p)$. Since the $(q,Q)$ representation connects to the phase-space $(q,p)$ picture through the partial Fourier transform
\begin{equation}
\label{eq:partialFT}
\begin{split}
  \psi(q,Q) &= \frac{1}{\sqrt{2\pi\hbar}}\!\int_{\mathbb{R}} e^{\,i Qp/\hbar}\,\Psi(q,p)\,dp, \\
  \Psi(q,p) &= \frac{1}{\sqrt{2\pi\hbar}}\!\int_{\mathbb{R}} e^{\,-i Qp/\hbar}\,\psi(q,Q)\,dQ,
\end{split}
\end{equation}
the momentum reversal $p \mapsto -p$ in phase space becomes the parity involution $Q \mapsto -Q$ in the dual representation. The boundary condition $\Psi(q_w,p) = \Psi(q_w,-p)$ is therefore equivalent to the $Q$-parity constraint
\begin{equation}
\label{eq:Q-parity}
  \psi(q_w,Q,t)=\psi(q_w,-Q,t),\qquad q_w\in\{0,L\}.
\end{equation}

This equivalence emerges naturally from the parity properties of the Fourier kernel. Assume $\Psi(q_w,p)$ is even in $p$: $\Psi(q_w,p) = \Psi(q_w,-p)$. Using Euler's formula $e^{iQp/\hbar} = \cos(Qp/\hbar) + i\sin(Qp/\hbar)$, the transform becomes
\begin{equation}
\begin{split}
\psi(q_w,Q) &= \frac{1}{\sqrt{2\pi\hbar}}\int_{\mathbb{R}} \Big[\cos(Qp/\hbar) + i\sin(Qp/\hbar)\Big]\Psi(q_w,p)\,dp.
\end{split}
\end{equation}
Since $\cos(Qp/\hbar)$ is even in $p$ and $\sin(Qp/\hbar)$ is odd in $p$, while $\Psi(q_w,p)$ is even in $p$, the sine integrand integrates to zero over $\mathbb{R}$. Only the cosine term survives:
\begin{equation}
\psi(q_w,Q) = \frac{1}{\sqrt{2\pi\hbar}}\int_{\mathbb{R}}\cos(Qp/\hbar)\,\Psi(q_w,p)\,dp.
\end{equation}
This expression is manifestly even in $Q$ since $\cos(Qp/\hbar) = \cos((-Q)p/\hbar)$. The converse follows by the same argument applied to the inverse transform: if $\psi(q_w,Q)$ is even in $Q$, then $\Psi(q_w,p)$ emerges as even in $p$.

The $Q$-parity condition \eqref{eq:Q-parity} has immediate physical implications for probability conservation.  
We first obtain a continuity equation by multiplying \eqref{eq:box-PDE} by $\psi^*$, subtracting the complex conjugate of this result, and identifying the resulting terms as divergences of probability currents.
Defining the density $\rho = |\psi|^2$ and using the commutativity 
$\partial_{qQ} = \partial_{Qq}$, one obtains
\begin{equation}
\label{eq:currents}
\begin{aligned}
  \partial_t \rho \;+\; \partial_q J_q \;+\; \partial_Q J_Q &= 0, \\[4pt]
  J_q &= \frac{\hbar}{m}\,\mathrm{Im}\!\left(\psi^*\,\partial_Q \psi\right), \\[4pt]
  J_Q &= \frac{\hbar}{m}\,\mathrm{Im}\!\left(\psi^*\,\partial_q \psi\right).
\end{aligned}
\end{equation}

Now at $q = q_w$, the parity condition \eqref{eq:Q-parity} makes $\psi$ even and $\partial_Q\psi$ odd in $Q$, hence $J_q(q_w,Q,t) \equiv 0$ pointwise. This provides the local, mathematical expression of the no-leak condition: no probability current penetrates the elastic wall, consistent with classical specular reflection.

The same boundary condition ensures that the mixed-derivative operator $-\partial_q\partial_Q$ admits a self-adjoint realization. Green's identity shows that for $\psi_{1,2}$ in the appropriate Sobolev space,
\begin{equation}
\label{eq:boundary-form}
\begin{split}
  &\langle\psi_1,-\partial_q\partial_Q\psi_2\rangle-\langle-\partial_q\partial_Q\psi_1,\psi_2\rangle \\
  &= \int_{\mathbb{R}}\! dQ\;\big[\psi_1^*(\partial_Q\psi_2)-(\partial_Q\psi_1)^*\psi_2\big]\Big|_{q=0}^{q=L}.
\end{split}
\end{equation}

The $Q$-parity boundary condition renders the integrand odd in $Q$, causing the boundary form to vanish and defining the self-adjoint operator
\begin{equation}
\begin{split}
  \widehat{L}_{\rm box} &:= -\frac{\hbar^2}{m}\,\partial_q\partial_Q, \\
  \mathcal{D}(\widehat{L}_{\rm box}) &= \{\psi\in H^1_{\rm loc}:\ \text{\eqref{eq:Q-parity} holds at}\ q=0,L\}.
\end{split}
\end{equation}

By Stone's theorem, $U(t) = e^{-it\widehat{L}_{\rm box}/\hbar}$ generates a one-parameter unitary group, ensuring unitary time evolution.

To find the energy spectrum, we seek separated solutions $\psi(q,Q,t) = \phi(q)\chi(Q)e^{-iEt/\hbar}$. Substituting into \eqref{eq:box-PDE} yields $\phi'(q)\chi'(Q) = -(mE/\hbar^2)\phi(q)\chi(Q)$. Plane wave solutions $\phi = e^{ikq}$, $\chi = e^{i\kappa Q}$ satisfy this with the dispersion relation
\begin{equation}
\label{eq:dispersion}
  E=\frac{\hbar^2}{m}\,k\,\kappa.
\end{equation}

Elastic reflection at the walls discretizes the solution in the $q$-direction: $k_n = n\pi/L$ with $n \in \mathbb{N}$.
leading to standing wave modes
\begin{equation}
  \phi_n(q)=\sin\frac{n\pi q}{L}\quad\text{or}\quad \phi_n(q)=\cos\frac{n\pi q}{L}.
\end{equation}

However, the $Q$-parity condition \eqref{eq:Q-parity} requires only that $\chi(Q)$ be even: $\chi_\kappa(Q) = \cos(\kappa Q)$ with $\kappa \in \mathbb{R}$ remaining continuous. Since $Q$ is unbounded, there is no mechanism to discretize $\kappa$. The spectrum therefore organizes into energy bands
\begin{equation}
\label{eq:band-spectrum}
\begin{split}
  E_{n,\kappa} &= \frac{\hbar^2}{m}\,\frac{n\pi}{L}\,\kappa, \\
  &\qquad n\in\mathbb{N},\ \kappa\in\mathbb{R}.
\end{split}
\end{equation}

This band structure embodies the central result: spatial confinement discretizes the mode index $n$ but leaves the energy continuous through the unbounded dual coordinate $Q$. In other words, the wall is at $q=0,L$, but the boundary condition it imposes is $\psi(q_w,Q)=\psi(q_w,-Q)$; this discretizes the $q$-modes (fixing $k$) while leaving the $Q$-label $\kappa$ continuous. Unlike the quantum case, where a single confined coordinate produces the discrete spectrum $E_n = n^2\pi^2\hbar^2/(2mL^2)$, the KvN formalism yields continuous energy within each spatial mode. 

The distinction lies in the operator structure and spatial dimensionality. In quantum mechanics, the time-independent Schr\"odinger equation $-\frac{\hbar^2}{2m}\partial_q^2\psi + V\psi = E\psi$ is a second-order elliptic PDE in the single confined coordinate $q \in [0,L]$. Boundary conditions lead to a discrete eigenvalue spectrum reflecting the compactness of the resolvent for elliptic operators on bounded domains. In KvN mechanics, the generator $-\frac{\hbar^2}{m}\partial_q\partial_Q$ is a first-order mixed-derivative operator acting on two coordinates, only one of which ($q$) is spatially confined. The physically appropriate $Q$-parity condition ensures self-adjointness and probability conservation, but cannot discretize energy because $Q$ remains unbounded.

This reveals that energy quantization requires confinement in \emph{all} dynamically active coordinates. Since classical phase space is intrinsically two-dimensional (per degree of freedom), and only position is confined by walls, the conjugate momentum coordinate retains its continuous character. The continuous spectrum \eqref{eq:band-spectrum} therefore faithfully represents the underlying classical mechanical content, where trajectories can have arbitrarily small momentum variations.

Geometrically, the evolution \eqref{eq:box-PDE} describes transport along straight-line characteristics in the $(q,Q)$ plane with $dQ/dq = \text{const}$ in the bulk. When a characteristic reaches $q = 0$ or $q = L$, it reflects according to $Q \mapsto -Q$, corresponding to the momentum reversal $p \mapsto -p$ of elastic collision. This preserves the classical phase space structure while maintaining unitary evolution.

The time evolution can be constructed rigorously using the free-space propagator. For $t \neq 0$, the unrestricted $(q,Q) \in \mathbb{R}^2$ evolution kernel is
\begin{equation}
\label{eq:free-kernel}
\begin{split}
  K_0(t;q,Q;q',Q') &= \frac{m}{2\pi\hbar t}\, \\
  &\quad \times \exp\!\left[\frac{i m}{\hbar t}\,(q-q')\,(Q-Q')\right].
\end{split}
\end{equation}

The confined propagator follows from the method of images, incorporating both spatial reflections and $Q$-parity:
\begin{equation}
\label{eq:box-kernel}
\begin{split}
  K_{\rm box}(t) &= \sum_{n\in\mathbb{Z}} \Big[K_0\big(t;q-2nL,\ Q;\ q',Q'\big) \\
  &\quad + K_0\big(t;q-(2nL-2q'),\ Q;\ q',-Q'\big)\Big].
\end{split}
\end{equation}

This construction satisfies the evolution equation and boundary conditions while preserving unitarity through orthogonality of the image contributions.

The analysis demonstrates that discretization is not an automatic consequence of Hilbert space boundary conditions, but depends critically on the specific operator structure and the dimensionality of confinement. Energy quantization in quantum mechanics arises through the non-commutativity $[\hat{q},\hat{p}] = i\hbar$, which allows a single confined coordinate to determine the full spectral structure. In KvN mechanics, where $[\hat{q},\hat{p}] = 0$, position and momentum coordinates retain their independent classical roles, and spatial confinement alone cannot discretize the energy spectrum.

\section{Conclusion and Outlook}

In this work we have revisited the paradigmatic problem of a particle in a box through the lens of Koopman--von Neumann mechanics. By imposing physically motivated $Q$-parity boundary conditions, we demonstrated that spatial confinement alone does not enforce energy quantization within the Hilbert-space formulation of classical dynamics. Instead of producing discrete energy levels as in quantum mechanics, the KvN spectrum separates into continuous energy bands. The discrete label $n$ arises from spatial confinement, $k_n = n\pi/L$, while the energy within each band varies continuously with the unbounded dual coordinate $Q$ (or its conjugate wave number $\kappa$). In this sense, $n$ indexes the bands, while $\kappa$ parametrizes the continuous energies within each band. The band here is not a quantum band, but simply a continuous family of energies labeled by the discrete mode number. This reflects the classical nature of KvN mechanics: since $[\hat q,\hat p]=0$, momentum is not quantized, so spatial confinement fixes $n$ while $\kappa$ remains continuous.


Our analysis clarifies common misconceptions regarding the role of boundary conditions in KvN mechanics, highlighting that the physically appropriate wall condition is elastic reflection rather than wavefunction vanishing. The latter would correspond, in classical terms, to an absorbing wall that removes trajectories, which contradicts the intended picture of perfectly reflecting, probability-conserving boundaries. By contrast, the reflection condition ensures both conservation of probability current and self-adjointness of the KvN generator, while faithfully reproducing classical mechanical content in a Hilbert-space framework.

\bibliography{ref}

\begin{thebibliography}{30}%
\makeatletter
\providecommand \@ifxundefined [1]{%
 \@ifx{#1\undefined}
}%
\providecommand \@ifnum [1]{%
 \ifnum #1\expandafter \@firstoftwo
 \else \expandafter \@secondoftwo
 \fi
}%
\providecommand \@ifx [1]{%
 \ifx #1\expandafter \@firstoftwo
 \else \expandafter \@secondoftwo
 \fi
}%
\providecommand \natexlab [1]{#1}%
\providecommand \enquote  [1]{``#1''}%
\providecommand \bibnamefont  [1]{#1}%
\providecommand \bibfnamefont [1]{#1}%
\providecommand \citenamefont [1]{#1}%
\providecommand \href@noop [0]{\@secondoftwo}%
\providecommand \href [0]{\begingroup \@sanitize@url \@href}%
\providecommand \@href[1]{\@@startlink{#1}\@@href}%
\providecommand \@@href[1]{\endgroup#1\@@endlink}%
\providecommand \@sanitize@url [0]{\catcode `\\12\catcode `\$12\catcode `\&12\catcode `\#12\catcode `\^12\catcode `\_12\catcode `\%12\relax}%
\providecommand \@@startlink[1]{}%
\providecommand \@@endlink[0]{}%
\providecommand \url  [0]{\begingroup\@sanitize@url \@url }%
\providecommand \@url [1]{\endgroup\@href {#1}{\urlprefix }}%
\providecommand \urlprefix  [0]{URL }%
\providecommand \Eprint [0]{\href }%
\providecommand \doibase [0]{https://doi.org/}%
\providecommand \selectlanguage [0]{\@gobble}%
\providecommand \bibinfo  [0]{\@secondoftwo}%
\providecommand \bibfield  [0]{\@secondoftwo}%
\providecommand \translation [1]{[#1]}%
\providecommand \BibitemOpen [0]{}%
\providecommand \bibitemStop [0]{}%
\providecommand \bibitemNoStop [0]{.\EOS\space}%
\providecommand \EOS [0]{\spacefactor3000\relax}%
\providecommand \BibitemShut  [1]{\csname bibitem#1\endcsname}%
\let\auto@bib@innerbib\@empty
\bibitem [{\citenamefont {Koopman}(1931)}]{9}%
  \BibitemOpen
  \bibfield  {author} {\bibinfo {author} {\bibfnamefont {B.~O.}\ \bibnamefont {Koopman}},\ }\bibfield  {title} {\bibinfo {title} {Hamiltonian {Systems} and {Transformation} in {Hilbert} {Space}},\ }\href {https://doi.org/10.1073/pnas.17.5.315} {\bibfield  {journal} {\bibinfo  {journal} {Proc. Natl. Acad. Sci. USA}\ }\textbf {\bibinfo {volume} {17}},\ \bibinfo {pages} {315} (\bibinfo {year} {1931})}\BibitemShut {NoStop}%
\bibitem [{\citenamefont {von Neumann}(1932)}]{8}%
  \BibitemOpen
  \bibfield  {author} {\bibinfo {author} {\bibfnamefont {J.}~\bibnamefont {von Neumann}},\ }\bibfield  {title} {\bibinfo {title} {Zur {Operatorenmethode} {In} {Der} {Klassischen} {Mechanik}},\ }\href {https://doi.org/10.2307/1968537} {\bibfield  {journal} {\bibinfo  {journal} {The Annals of Mathematics}\ }\textbf {\bibinfo {volume} {33}},\ \bibinfo {pages} {587} (\bibinfo {year} {1932})}\BibitemShut {NoStop}%
\bibitem [{\citenamefont {Sen}\ and\ \citenamefont {Silagadze}(2020)}]{Sen2020}%
  \BibitemOpen
  \bibfield  {author} {\bibinfo {author} {\bibfnamefont {A.}~\bibnamefont {Sen}}\ and\ \bibinfo {author} {\bibfnamefont {Z.}~\bibnamefont {Silagadze}},\ }\bibfield  {title} {\bibinfo {title} {{Ermakov-Lewis} invariant in {Koopman-von Neumann} mechanics},\ }\href {https://doi.org/10.1007/s10773-020-04492-3} {\bibfield  {journal} {\bibinfo  {journal} {International Journal of Theoretical Physics}\ }\textbf {\bibinfo {volume} {59}},\ \bibinfo {pages} {2187–2190} (\bibinfo {year} {2020})}\BibitemShut {NoStop}%
\bibitem [{\citenamefont {Sen}\ \emph {et~al.}(2023)\citenamefont {Sen}, \citenamefont {Parida}, \citenamefont {Dhasmana},\ and\ \citenamefont {Silagadze}}]{Sen2023}%
  \BibitemOpen
  \bibfield  {author} {\bibinfo {author} {\bibfnamefont {A.}~\bibnamefont {Sen}}, \bibinfo {author} {\bibfnamefont {B.~K.}\ \bibnamefont {Parida}}, \bibinfo {author} {\bibfnamefont {S.}~\bibnamefont {Dhasmana}},\ and\ \bibinfo {author} {\bibfnamefont {Z.~K.}\ \bibnamefont {Silagadze}},\ }\bibfield  {title} {\bibinfo {title} {Eisenhart lift of {Koopman-von Neumann} mechanics},\ }\href {https://doi.org/10.1016/j.geomphys.2022.104732} {\bibfield  {journal} {\bibinfo  {journal} {Journal of Geometry and Physics}\ }\textbf {\bibinfo {volume} {185}},\ \bibinfo {pages} {104732} (\bibinfo {year} {2023})}\BibitemShut {NoStop}%
\bibitem [{\citenamefont {Parida}\ \emph {et~al.}(2023)\citenamefont {Parida}, \citenamefont {Sen}, \citenamefont {Dhasmana},\ and\ \citenamefont {Silagadze}}]{Sen_2023}%
  \BibitemOpen
  \bibfield  {author} {\bibinfo {author} {\bibfnamefont {B.~K.}\ \bibnamefont {Parida}}, \bibinfo {author} {\bibfnamefont {A.}~\bibnamefont {Sen}}, \bibinfo {author} {\bibfnamefont {S.}~\bibnamefont {Dhasmana}},\ and\ \bibinfo {author} {\bibfnamefont {Z.~K.}\ \bibnamefont {Silagadze}},\ }\bibfield  {title} {\bibinfo {title} {{Lévy-Leblond} equation and {Eisenhart–Duval} lift in {Koopman–von Neumann} mechanics},\ }\bibfield  {journal} {\bibinfo  {journal} {Modern Physics Letters A}\ }\textbf {\bibinfo {volume} {38}},\ \href {https://doi.org/10.1142/s0217732323501493} {10.1142/s0217732323501493} (\bibinfo {year} {2023})\BibitemShut {NoStop}%
\bibitem [{\citenamefont {Chashchina}\ \emph {et~al.}(2020)\citenamefont {Chashchina}, \citenamefont {Sen},\ and\ \citenamefont {Silagadze}}]{Sen-2023}%
  \BibitemOpen
  \bibfield  {author} {\bibinfo {author} {\bibfnamefont {O.~I.}\ \bibnamefont {Chashchina}}, \bibinfo {author} {\bibfnamefont {A.}~\bibnamefont {Sen}},\ and\ \bibinfo {author} {\bibfnamefont {Z.~K.}\ \bibnamefont {Silagadze}},\ }\bibfield  {title} {\bibinfo {title} {On deformations of classical mechanics due to {Planck-scale} physics},\ }\href {https://doi.org/10.1142/s0218271820500704} {\bibfield  {journal} {\bibinfo  {journal} {International Journal of Modern Physics D}\ }\textbf {\bibinfo {volume} {29}},\ \bibinfo {pages} {2050070} (\bibinfo {year} {2020})}\BibitemShut {NoStop}%
\bibitem [{\citenamefont {Bondar}\ \emph {et~al.}(2012)\citenamefont {Bondar}, \citenamefont {Cabrera}, \citenamefont {Lompay}, \citenamefont {Ivanov},\ and\ \citenamefont {Rabitz}}]{Bondar_2012a}%
  \BibitemOpen
  \bibfield  {author} {\bibinfo {author} {\bibfnamefont {D.~I.}\ \bibnamefont {Bondar}}, \bibinfo {author} {\bibfnamefont {R.}~\bibnamefont {Cabrera}}, \bibinfo {author} {\bibfnamefont {R.~R.}\ \bibnamefont {Lompay}}, \bibinfo {author} {\bibfnamefont {M.~Y.}\ \bibnamefont {Ivanov}},\ and\ \bibinfo {author} {\bibfnamefont {H.~A.}\ \bibnamefont {Rabitz}},\ }\bibfield  {title} {\bibinfo {title} {Operational dynamic modeling transcending quantum and classical mechanics},\ }\href {https://doi.org/10.1103/PhysRevLett.109.190403} {\bibfield  {journal} {\bibinfo  {journal} {Phys. Rev. Lett.}\ }\textbf {\bibinfo {volume} {109}},\ \bibinfo {pages} {190403} (\bibinfo {year} {2012})}\BibitemShut {NoStop}%
\bibitem [{\citenamefont {MAURO}(2002)}]{MAURO2002}%
  \BibitemOpen
  \bibfield  {author} {\bibinfo {author} {\bibfnamefont {D.}~\bibnamefont {MAURO}},\ }\bibfield  {title} {\bibinfo {title} {On koopman–von neumann waves},\ }\href {https://doi.org/10.1142/s0217751x02009680} {\bibfield  {journal} {\bibinfo  {journal} {International Journal of Modern Physics A}\ }\textbf {\bibinfo {volume} {17}},\ \bibinfo {pages} {1301–1325} (\bibinfo {year} {2002})}\BibitemShut {NoStop}%
\bibitem [{\citenamefont {GOZZI}\ and\ \citenamefont {MAURO}(2004)}]{GOZZI2004}%
  \BibitemOpen
  \bibfield  {author} {\bibinfo {author} {\bibfnamefont {E.}~\bibnamefont {GOZZI}}\ and\ \bibinfo {author} {\bibfnamefont {D.}~\bibnamefont {MAURO}},\ }\bibfield  {title} {\bibinfo {title} {On koopman–von neumann waves ii},\ }\href {https://doi.org/10.1142/s0217751x04017872} {\bibfield  {journal} {\bibinfo  {journal} {International Journal of Modern Physics A}\ }\textbf {\bibinfo {volume} {19}},\ \bibinfo {pages} {1475–1493} (\bibinfo {year} {2004})}\BibitemShut {NoStop}%
\bibitem [{\citenamefont {Klein}(2017)}]{Klein2017}%
  \BibitemOpen
  \bibfield  {author} {\bibinfo {author} {\bibfnamefont {U.}~\bibnamefont {Klein}},\ }\bibfield  {title} {\bibinfo {title} {From koopman–von neumann theory to quantum theory},\ }\href {https://doi.org/10.1007/s40509-017-0113-2} {\bibfield  {journal} {\bibinfo  {journal} {Quantum Studies: Mathematics and Foundations}\ }\textbf {\bibinfo {volume} {5}},\ \bibinfo {pages} {219–227} (\bibinfo {year} {2017})}\BibitemShut {NoStop}%
\bibitem [{\citenamefont {Joseph}(2020)}]{Joseph2020}%
  \BibitemOpen
  \bibfield  {author} {\bibinfo {author} {\bibfnamefont {I.}~\bibnamefont {Joseph}},\ }\bibfield  {title} {\bibinfo {title} {Koopman–von neumann approach to quantum simulation of nonlinear classical dynamics},\ }\bibfield  {journal} {\bibinfo  {journal} {Physical Review Research}\ }\textbf {\bibinfo {volume} {2}},\ \href {https://doi.org/10.1103/physrevresearch.2.043102} {10.1103/physrevresearch.2.043102} (\bibinfo {year} {2020})\BibitemShut {NoStop}%
\bibitem [{\citenamefont {Deotto}\ \emph {et~al.}(2003{\natexlab{a}})\citenamefont {Deotto}, \citenamefont {Gozzi},\ and\ \citenamefont {Mauro}}]{Deotto2003}%
  \BibitemOpen
  \bibfield  {author} {\bibinfo {author} {\bibfnamefont {E.}~\bibnamefont {Deotto}}, \bibinfo {author} {\bibfnamefont {E.}~\bibnamefont {Gozzi}},\ and\ \bibinfo {author} {\bibfnamefont {D.}~\bibnamefont {Mauro}},\ }\bibfield  {title} {\bibinfo {title} {Hilbert space structure in classical mechanics. i},\ }\href {https://doi.org/10.1063/1.1623333} {\bibfield  {journal} {\bibinfo  {journal} {Journal of Mathematical Physics}\ }\textbf {\bibinfo {volume} {44}},\ \bibinfo {pages} {5902–5936} (\bibinfo {year} {2003}{\natexlab{a}})}\BibitemShut {NoStop}%
\bibitem [{\citenamefont {Deotto}\ \emph {et~al.}(2003{\natexlab{b}})\citenamefont {Deotto}, \citenamefont {Gozzi},\ and\ \citenamefont {Mauro}}]{Deotto22003}%
  \BibitemOpen
  \bibfield  {author} {\bibinfo {author} {\bibfnamefont {E.}~\bibnamefont {Deotto}}, \bibinfo {author} {\bibfnamefont {E.}~\bibnamefont {Gozzi}},\ and\ \bibinfo {author} {\bibfnamefont {D.}~\bibnamefont {Mauro}},\ }\bibfield  {title} {\bibinfo {title} {Hilbert space structure in classical mechanics. ii},\ }\href {https://doi.org/10.1063/1.1623334} {\bibfield  {journal} {\bibinfo  {journal} {Journal of Mathematical Physics}\ }\textbf {\bibinfo {volume} {44}},\ \bibinfo {pages} {5937–5957} (\bibinfo {year} {2003}{\natexlab{b}})}\BibitemShut {NoStop}%
\bibitem [{\citenamefont {Mauro}(2002)}]{Mauro:2001rm}%
  \BibitemOpen
  \bibfield  {author} {\bibinfo {author} {\bibfnamefont {D.}~\bibnamefont {Mauro}},\ }\bibfield  {title} {\bibinfo {title} {{On {Koopman-von Neumann} waves}},\ }\href {https://doi.org/10.1142/S0217751X02009680} {\bibfield  {journal} {\bibinfo  {journal} {Int. J. Mod. Phys. A}\ }\textbf {\bibinfo {volume} {17}},\ \bibinfo {pages} {1301} (\bibinfo {year} {2002})},\ \Eprint {https://arxiv.org/abs/quant-ph/0105112} {arXiv:quant-ph/0105112} \BibitemShut {NoStop}%
\bibitem [{\citenamefont {Mauro}(2003)}]{Mauro}%
  \BibitemOpen
  \bibfield  {author} {\bibinfo {author} {\bibfnamefont {D.}~\bibnamefont {Mauro}},\ }\href {https://doi.org/10.48550/ARXIV.QUANT-PH/0301172} {\bibinfo {title} {Topics in {Koopman-von Neumann} theory}} (\bibinfo {year} {2003})\BibitemShut {NoStop}%
\bibitem [{\citenamefont {Shankar}(1994)}]{Shankar_1994}%
  \BibitemOpen
  \bibfield  {author} {\bibinfo {author} {\bibfnamefont {R.}~\bibnamefont {Shankar}},\ }\href {https://doi.org/10.1007/978-1-4757-0576-8} {\emph {\bibinfo {title} {Principles of Quantum Mechanics}}}\ (\bibinfo  {publisher} {Springer {US}},\ \bibinfo {address} {New York},\ \bibinfo {year} {1994})\BibitemShut {NoStop}%
\bibitem [{\citenamefont {Klein}(2018)}]{Klein_2018}%
  \BibitemOpen
  \bibfield  {author} {\bibinfo {author} {\bibfnamefont {U.}~\bibnamefont {Klein}},\ }\bibfield  {title} {\bibinfo {title} {From {Koopman–von Neumann} theory to quantum theory},\ }\href {https://doi.org/10.1007/s40509-017-0113-2} {\bibfield  {journal} {\bibinfo  {journal} {Quantum Stud.: Math. Found.}\ }\textbf {\bibinfo {volume} {5}},\ \bibinfo {pages} {219} (\bibinfo {year} {2018})}\BibitemShut {NoStop}%
\bibitem [{\citenamefont {Giulini}\ \emph {et~al.}(1996)\citenamefont {Giulini}, \citenamefont {Kiefer}, \citenamefont {Joos}, \citenamefont {Kupsch}, \citenamefont {Stamatescu},\ and\ \citenamefont {Zeh}}]{Giulini:1996nw}%
  \BibitemOpen
  \bibfield  {author} {\bibinfo {author} {\bibfnamefont {D.}~\bibnamefont {Giulini}}, \bibinfo {author} {\bibfnamefont {C.}~\bibnamefont {Kiefer}}, \bibinfo {author} {\bibfnamefont {E.}~\bibnamefont {Joos}}, \bibinfo {author} {\bibfnamefont {J.}~\bibnamefont {Kupsch}}, \bibinfo {author} {\bibfnamefont {I.~O.}\ \bibnamefont {Stamatescu}},\ and\ \bibinfo {author} {\bibfnamefont {H.~D.}\ \bibnamefont {Zeh}},\ }\href {https://doi.org/10.1007/978-3-662-05328-7} {\emph {\bibinfo {title} {{Decoherence and the appearance of a classical world in quantum theory}}}}\ (\bibinfo  {publisher} {Springer},\ \bibinfo {address} {Berlin},\ \bibinfo {year} {1996})\BibitemShut {NoStop}%
\bibitem [{\citenamefont {Klein}(2012)}]{Klein2012}%
  \BibitemOpen
  \bibfield  {author} {\bibinfo {author} {\bibfnamefont {U.}~\bibnamefont {Klein}},\ }\bibfield  {title} {\bibinfo {title} {What is the limit $\hslash${\hspace{0.167em}}$\rightarrow${\hspace{0.167em}}0 of quantum theory?},\ }\href {https://doi.org/10.1119/1.4751274} {\bibfield  {journal} {\bibinfo  {journal} {Am. J. Phys.}\ }\textbf {\bibinfo {volume} {80}},\ \bibinfo {pages} {1009} (\bibinfo {year} {2012})}\BibitemShut {NoStop}%
\bibitem [{\citenamefont {Gozzi}\ and\ \citenamefont {Mauro}(2004)}]{Gozzi:2003sh}%
  \BibitemOpen
  \bibfield  {author} {\bibinfo {author} {\bibfnamefont {E.}~\bibnamefont {Gozzi}}\ and\ \bibinfo {author} {\bibfnamefont {D.}~\bibnamefont {Mauro}},\ }\bibfield  {title} {\bibinfo {title} {{On {Koopman-von Neumann} waves {II}}},\ }\href {https://doi.org/10.1142/S0217751X04017872} {\bibfield  {journal} {\bibinfo  {journal} {Int. J. Mod. Phys. A}\ }\textbf {\bibinfo {volume} {19}},\ \bibinfo {pages} {1475} (\bibinfo {year} {2004})},\ \Eprint {https://arxiv.org/abs/quant-ph/0306029} {arXiv:quant-ph/0306029} \BibitemShut {NoStop}%
\bibitem [{\citenamefont {Wilczek}(2022)}]{wilczek_2022}%
  \BibitemOpen
  \bibfield  {author} {\bibinfo {author} {\bibfnamefont {F.}~\bibnamefont {Wilczek}},\ }\href {http://frankwilczek.com/2015/koopmanVonNeumann02.pdf} {\bibinfo {title} {Notes on {Koopman von Neumann} mechanics, and a step beyond}} (\bibinfo {year} {2022})\BibitemShut {NoStop}%
\bibitem [{\citenamefont {Sudarshan}(1976)}]{Sudarshan_1976}%
  \BibitemOpen
  \bibfield  {author} {\bibinfo {author} {\bibfnamefont {E.~C.~G.}\ \bibnamefont {Sudarshan}},\ }\bibfield  {title} {\bibinfo {title} {Interaction between classical and quantum systems and the measurement of quantum observables},\ }\href {https://doi.org/10.1007/BF02847120} {\bibfield  {journal} {\bibinfo  {journal} {Pramana}\ }\textbf {\bibinfo {volume} {6}},\ \bibinfo {pages} {117} (\bibinfo {year} {1976})}\BibitemShut {NoStop}%
\bibitem [{\citenamefont {Sherry}\ and\ \citenamefont {Sudarshan}(1978)}]{Sherry1978}%
  \BibitemOpen
  \bibfield  {author} {\bibinfo {author} {\bibfnamefont {T.~N.}\ \bibnamefont {Sherry}}\ and\ \bibinfo {author} {\bibfnamefont {E.~C.~G.}\ \bibnamefont {Sudarshan}},\ }\bibfield  {title} {\bibinfo {title} {Interaction between classical and quantum systems: A new approach to quantum measurement.{I}},\ }\href {https://doi.org/10.1103/physrevd.18.4580} {\bibfield  {journal} {\bibinfo  {journal} {Phys. Rev. D}\ }\textbf {\bibinfo {volume} {18}},\ \bibinfo {pages} {4580} (\bibinfo {year} {1978})}\BibitemShut {NoStop}%
\bibitem [{\citenamefont {Bondar}\ \emph {et~al.}(2013)\citenamefont {Bondar}, \citenamefont {Cabrera}, \citenamefont {Zhdanov},\ and\ \citenamefont {Rabitz}}]{Bondar_2013}%
  \BibitemOpen
  \bibfield  {author} {\bibinfo {author} {\bibfnamefont {D.~I.}\ \bibnamefont {Bondar}}, \bibinfo {author} {\bibfnamefont {R.}~\bibnamefont {Cabrera}}, \bibinfo {author} {\bibfnamefont {D.~V.}\ \bibnamefont {Zhdanov}},\ and\ \bibinfo {author} {\bibfnamefont {H.~A.}\ \bibnamefont {Rabitz}},\ }\bibfield  {title} {\bibinfo {title} {Wigner phase-space distribution as a wave function},\ }\href {https://doi.org/10.1103/physreva.88.052108} {\bibfield  {journal} {\bibinfo  {journal} {Phys. Rev. A}\ }\textbf {\bibinfo {volume} {88}},\ \bibinfo {pages} {052108} (\bibinfo {year} {2013})}\BibitemShut {NoStop}%
\bibitem [{\citenamefont {Hillery}\ \emph {et~al.}(1984)\citenamefont {Hillery}, \citenamefont {O'Connell}, \citenamefont {Scully},\ and\ \citenamefont {Wigner}}]{Hillery1984}%
  \BibitemOpen
  \bibfield  {author} {\bibinfo {author} {\bibfnamefont {M.}~\bibnamefont {Hillery}}, \bibinfo {author} {\bibfnamefont {R.}~\bibnamefont {O'Connell}}, \bibinfo {author} {\bibfnamefont {M.}~\bibnamefont {Scully}},\ and\ \bibinfo {author} {\bibfnamefont {E.}~\bibnamefont {Wigner}},\ }\bibfield  {title} {\bibinfo {title} {Distribution functions in physics: Fundamentals},\ }\href {https://doi.org/10.1016/0370-1573(84)90160-1} {\bibfield  {journal} {\bibinfo  {journal} {Phys. Rep.}\ }\textbf {\bibinfo {volume} {106}},\ \bibinfo {pages} {121} (\bibinfo {year} {1984})}\BibitemShut {NoStop}%
\bibitem [{\citenamefont {Gozzi}\ and\ \citenamefont {Mauro}(2002)}]{Gozzi:2001he}%
  \BibitemOpen
  \bibfield  {author} {\bibinfo {author} {\bibfnamefont {E.}~\bibnamefont {Gozzi}}\ and\ \bibinfo {author} {\bibfnamefont {D.}~\bibnamefont {Mauro}},\ }\bibfield  {title} {\bibinfo {title} {{Minimal coupling in {Koopman-von Neumann} theory}},\ }\href {https://doi.org/10.1006/aphy.2001.6206} {\bibfield  {journal} {\bibinfo  {journal} {Annals Phys.}\ }\textbf {\bibinfo {volume} {296}},\ \bibinfo {pages} {152} (\bibinfo {year} {2002})},\ \Eprint {https://arxiv.org/abs/quant-ph/0105113} {arXiv:quant-ph/0105113} \BibitemShut {NoStop}%
\bibitem [{\citenamefont {Abrikosov}\ \emph {et~al.}(2005)\citenamefont {Abrikosov}, \citenamefont {Gozzi},\ and\ \citenamefont {Mauro}}]{Abrikosov:2004cf}%
  \BibitemOpen
  \bibfield  {author} {\bibinfo {author} {\bibfnamefont {A.~A.}\ \bibnamefont {Abrikosov}, \bibfnamefont {Jr.}}, \bibinfo {author} {\bibfnamefont {E.}~\bibnamefont {Gozzi}},\ and\ \bibinfo {author} {\bibfnamefont {D.}~\bibnamefont {Mauro}},\ }\bibfield  {title} {\bibinfo {title} {{Geometric dequantization}},\ }\href {https://doi.org/10.1016/j.aop.2004.12.001} {\bibfield  {journal} {\bibinfo  {journal} {Annals Phys.}\ }\textbf {\bibinfo {volume} {317}},\ \bibinfo {pages} {24} (\bibinfo {year} {2005})},\ \Eprint {https://arxiv.org/abs/quant-ph/0406028} {arXiv:quant-ph/0406028} \BibitemShut {NoStop}%
\bibitem [{\citenamefont {Padmanabhan}(2015)}]{Padmanabhan2015}%
  \BibitemOpen
  \bibfield  {author} {\bibinfo {author} {\bibfnamefont {T.}~\bibnamefont {Padmanabhan}},\ }\href {https://doi.org/10.1007/978-3-319-13443-7} {\emph {\bibinfo {title} {Sleeping Beauties in Theoretical Physics: 26 Surprising Insights}}}\ (\bibinfo  {publisher} {Springer International Publishing},\ \bibinfo {year} {2015})\BibitemShut {NoStop}%
\bibitem [{\citenamefont {Sen}\ \emph {et~al.}(2020)\citenamefont {Sen}, \citenamefont {Dhasmana},\ and\ \citenamefont {Silagadze}}]{Sen2020free}%
  \BibitemOpen
  \bibfield  {author} {\bibinfo {author} {\bibfnamefont {A.}~\bibnamefont {Sen}}, \bibinfo {author} {\bibfnamefont {S.}~\bibnamefont {Dhasmana}},\ and\ \bibinfo {author} {\bibfnamefont {Z.~K.}\ \bibnamefont {Silagadze}},\ }\bibfield  {title} {\bibinfo {title} {Free fall in {KvN} mechanics and {Einstein’s} principle of equivalence},\ }\href {https://doi.org/10.1016/j.aop.2020.168302} {\bibfield  {journal} {\bibinfo  {journal} {Annals of Physics}\ }\textbf {\bibinfo {volume} {422}},\ \bibinfo {pages} {168302} (\bibinfo {year} {2020})}\BibitemShut {NoStop}%
\bibitem [{\citenamefont {Silagadze}(2023)}]{Silagadze2023}%
  \BibitemOpen
  \bibfield  {author} {\bibinfo {author} {\bibfnamefont {Z.~K.}\ \bibnamefont {Silagadze}},\ }\bibfield  {title} {\bibinfo {title} {Evading quantum mechanics à la {Sudarshan}: Quantum-mechanics-free subsystem as a realization of {Koopman-von Neumann} mechanics},\ }\bibfield  {journal} {\bibinfo  {journal} {Foundations of Physics}\ }\textbf {\bibinfo {volume} {53}},\ \href {https://doi.org/10.1007/s10701-023-00734-6} {10.1007/s10701-023-00734-6} (\bibinfo {year} {2023})\BibitemShut {NoStop}%
\end{thebibliography}%

\end{document}